\documentclass[12pt]{article}
\usepackage{amsmath,amsfonts,amssymb}
\usepackage[nosort]{cite}
\usepackage{hyperref}
\usepackage{graphicx}
\usepackage{color}
\usepackage{epsfig}
\usepackage{psfrag}	
\usepackage{tikz}
\usetikzlibrary{snakes}
\usepackage{slashed}
\usepackage{tensor}
\usepackage{ulem}
\usepackage[font=small,labelfont=bf,width=1\textwidth]{caption}
\usepackage{comment}
\usepackage[all]{xy}
\usepackage{multirow}
\usepackage{float}

\usepackage{color}

\usepackage{tocloft} 
\usepackage{ulem}
\usepackage{soul,xcolor}
\setstcolor{red}

\numberwithin{equation}{section}

\DeclareMathOperator{\Tr}{Tr}
\DeclareMathOperator{\tr}{tr}

\DeclareMathOperator{\diag}{diag}

\newcommand{\bpm}{\begin{pmatrix}}
\newcommand{\epm}{\end{pmatrix}}
\newcommand{\bbm}{\begin{bmatrix}}
\newcommand{\ebm}{\end{bmatrix}}

\def\bC {\mathbb{C}}

\def\bR {\mathbb{R}}

\def\bZ {\mathbb{Z}}

\newcommand{\bea}{\begin{eqnarray}}
\newcommand{\eea}{\end{eqnarray}}
\newcommand{\beq}{\begin{equation}}
\newcommand{\eeq}{\end{equation}}
\newcommand{\bal}{\begin{equation}\begin{aligned}}
\newcommand{\eal}{\end{aligned} \end{equation}}

\newcommand{\address}[1]{\vbox{\center\em#1}}
\renewcommand{\title}[1]{\vbox{\center\huge{#1}}\vspace{5mm}}

\newcommand{\cL}{{\mathcal L}}

\newcommand{\cN}{{\mathcal N}}
\newcommand{\cP}{{\mathcal P}}
\newcommand{\cQ}{{\mathcal Q}}

\begin{document}

\setstcolor{red}

\begin{titlepage}
\begin{center}

\vspace*{10mm}

\title{Classifying BPS bosonic Wilson loops in \\3d ${\cal N}=4$ Chern-Simons-matter theories}

\vspace{7mm}

\renewcommand{\thefootnote}{$\alph{footnote}$}

Nadav Drukker,%
\textsuperscript{1}
Ziwen Kong,%
\textsuperscript{1}
Malte Probst,%
\textsuperscript{1}
\\
Marcia Tenser,%
\textsuperscript{2}
and
Diego Trancanelli%
\textsuperscript{3,4}

\vskip 2mm
\address{
\textsuperscript{1}%
Department of Mathematics, King's College London,
\\
The Strand, WC2R 2LS London, United-Kingdom}

\address{
\textsuperscript{2}%
Dipartimento di Fisica, Universit\`a degli Studi di Milano-Bicocca,
 \\ \& \\
INFN Sezione di Milano-Bicocca, Piazza della Scienza 3, I-20126 Milano, Italy}

\address{
\textsuperscript{3}%
Institute of Physics, University of S\~ao Paulo,
05314-970 S\~ao Paulo, Brazil
}

\address{
\textsuperscript{4}%
 Dipartimento di Scienze Fisiche, Informatiche e Matematiche, \\
Universit\`a di Modena e Reggio Emilia, via G. Campi 213/A, 41125 Modena, Italy \\ \& \\
INFN Sezione di Bologna, via Irnerio 46, 40126 Bologna, Italy}

\vskip .5cm
{\tt \{nadav.drukker, mltprbst, marciatenser, dtrancan\}@gmail.com\\ ziwen.kong@kcl.ac.uk}

\renewcommand{\thefootnote}{\arabic{footnote}}
\setcounter{footnote}{0}

\end{center}
\vspace{4mm}
\abstract{
\normalsize{
\noindent
We study the possible BPS Wilson loops in three-dimensional ${\cal N}=4$ 
Chern-Simons-matter theory which involve only the gauge field and bilinears of the scalars. 
Previously known examples are the analogues of the Gaiotto-Yin loops preserving four 
supercharges and ``latitude'' loops preserving two. We carry out a careful classification and 
find, in addition, loops preserving three supercharges, further inequivalent classes of loops 
preserving two supercharges and loops preserving a single supercharge. For each of the 
classes of loops, we  present a representative example and analyse their full orbit under the 
broken symmetries.
}}
\vfill

\end{titlepage}

\section{Introduction and conclusions}
\label{sec:intro}

Over the past few years, more and more examples of supersymmetry preserving (BPS) line operators have been 
found~\cite{Drukker:2008zx, Chen:2008bp, Rey:2008bh, Drukker:2008jm, Ouyang:2015qma, 
Cooke:2015ila, Ouyang:2015iza, Ouyang:2015bmy,Mauri:2017whf, 
Mauri:2018fsf, drukker2020bps, Drukker:2020dvr, Drukker:2022ywj} 
in Chern-Simons-matter theories like ABJM~\cite{Aharony:2008ug}. For a relatively recent introduction to the topic, 
see~\cite{Drukker:2019bev}. While many papers discuss the bosonic 1/6 BPS loop and fermionic 1/2 BPS loop, 
there are in fact many more Wilson loops including rich moduli spaces of 1/6 BPS loops with fermionic 
fields and Wilson loops preserving fewer supercharges.

Following work on $\cN=2$ theories~\cite{drukker2020bps}, 
the recent papers~\cite{Drukker:2020dvr, Drukker:2022ywj} started to methodically treat the space of 
BPS Wilson loops in the context of theories with $\cN=4$ supersymmetry, called there ``hyperloops''. 
In the course of writing  the last paper in the 
series, we realised that even the case of Wilson loops with a single gauge field and no fermi fields is very 
rich and includes previously unnoticed operators preserving 1, 2 and 3 supercharges. As the classification of 
those was beyond the scope of that paper and requires many different tools, it is the topic of this paper.

A generic $\cN=4$ Chern-Simons-matter theory has vector multiplets, hypermultiplets as well as twisted 
hypers which can be organised graphically in either a circular or linear quiver 
diagram~\cite{Gaiotto:2008sd,Imamura:2008dt,Hosomichi:2008jd,Hama:2010av}. 
Restricting to bosonic fields, those are the vector fields $A_\mu$ and the bi-fundamental 
scalars in the hypermultiplet, $q^a$, $\bar q_a$ and in the twisted hypermultiplets, 
$\tilde q_{\dot a}$, $\bar{\tilde q}^{\dot a}$.

The scalar fields in the hypermultiplet have undotted indices and are doublets of the $SU(2)_L$ R-symmetry. 
The fermions $\psi_{\dot b}$, $\bar\psi^{\dot a}$ 
with dotted indices are charged instead under $SU(2)_R$. This is reversed 
for the twisted hypermultiplets. We use the usual epsilon symbols to raise and 
lower indices: $v^a=\epsilon^{ab}v_ b$ and $v_a=\epsilon_{ab}v^b$ with $\epsilon^{12}=\epsilon_{21}=1$, 
and likewise for the dotted indices.

We consider the theory on $S^3$ with the loops supported along the equator of this sphere 
with coordinate $\varphi$. The theories are conformal and this setup allows for conformal line 
operators preserving $SO(2,1)\times SO(2)\subset SO(4,1)$, but the 
loops we study are not necessarily conformal, as is discussed below. Still, we restrict the 
connection $\cL$ to have canonical dimension one, so it does not have dimensionful couplings. 
With this restriction, $\cL$ is comprised of the gauge field and bilinear of the scalars. 
This leads to the natural ansatz
\beq
\label{eqn:cL}
W=\Tr\cP\exp \oint i\cL\,d\varphi\,,
\qquad
\cL = A_\varphi + \frac{i}{k} q^a M\indices{_a^b} \bar q_b + 
\frac{i}{k} {\bar{\tilde q}}^{\dot a} \widetilde M\indices{_{\dot a}^{\dot b}} \tilde q_{\dot b}\,.
\eeq
$M$ and $\widetilde{M}$---the couplings of the scalar bilinears---are the main protagonists of this paper. 
We allow for them to have explicit $\varphi$-dependence, which breaks rotational symmetry and 
therefore also the conformal symmetry along the circle.

The scalar fields may be charged under a flavour group or in fact other gauge groups. The ansatz above 
assumes that they form singlets of those groups, since the supersymmetry variation \eqref{susy-var} 
below requires a cancellation between the variation of the gauge fields that are not charged under these other 
groups and that of the scalar bilinears.

More general Wilson loops in these theories include also fermi fields and a connection that is naturally 
extended to a supermatrix with multiple gauge fields, fermions and scalar bilinears. The restriction 
to bosonic structures still allows the scalar bilinears. Those appearing in \eqref{eqn:cL} are in the 
adjoint of the gauge group (plus the singlet), but in a quiver theory one could also construct 
bilinears of scalars from different multiplets $q^a\tilde q_{\dot a}$ which transform in the 
bifundamental of next to nearest neighbours in the quiver. Wilson loops without fermions but 
with these couplings are BPS only when the supercharge annihilates these bilinears. As this 
is a rather trivial additional constraint on our general analysis of \eqref{eqn:cL}, we do not 
discuss this possibility further.

With the increasingly rich and intricate structure of BPS loops in 3d theories, it is worth mentioning 
some of the possible applications of such operators. First, BPS protected quantities serve as a rich 
laboratory for refining the tools of quantum field theory, for example Seiberg-Witten 
theory~\cite{Seiberg:1994rs, Seiberg:1994aj} or $AdS$/CFT~\cite{Lee:1998bxa}. The circular 
BPS Wilson loop, in particular, is amenable to exact 
calculations~\cite{erickson, gross, Pestun:2009nn, Kapustin:2009kz, Marino:2009jd, Drukker:2010nc, Bianchi:2018bke} 
and the rich spectrum of BPS Wilson loops in 
4d~\cite{Zarembo:2002an, Drukker:2007dw, Drukker:2007yx, Drukker:2007qr, Dymarsky:2009si, Pestun:2009nn} 
allows for further exact results in quantities such as the bremsstrahlung 
function~\cite{correa, Fiol:2012sg} and its 3d 
generalisations~\cite{Cardinali:2012ru, Griguolo:2012iq, Correa:2014aga, Bianchi:2014laa, Bianchi:2017svd, 
Bianchi:2017ozk, Bianchi:2018scb}. Some of these loops or close analogues of them arise in our 
analysis and we hope that the new examples we uncover here will play a similar role in 
future work.

With this analysis of the bosonic loops complete, it is evident that the story of Wilson loops involving 
fermi fields and more than one gauge field is still richer than all those already identified 
in~\cite{Ouyang:2015qma, Cooke:2015ila, Ouyang:2015iza, Ouyang:2015bmy,Mauri:2017whf, 
Mauri:2018fsf, drukker2020bps, Drukker:2020dvr, Drukker:2022ywj}. This will be explored in the 
forthcoming paper~\cite{inprogress}.

\section{General analysis}
\label{sec:analysis}

We start by looking at the variation of \eqref{eqn:cL} under a generic supercharge. This leads to a set of conditions 
on the supercharges that can preserve such a loop. These are then used in the subsequent section to reconstruct 
the loop operator invariant under the possible supercharges.

${\cal N}=4$ Chern-Simons matter theories were constructed 
in~\cite{Gaiotto:2008sd,Imamura:2008dt,Hosomichi:2008jd,Hama:2010av} 
and the supersymmetry transformations in flat space were presented there. 
Those were adapted to $S^3$ in~\cite{Drukker:2020dvr}, relying also on 
the decomposition to $\cN=2$ theories and the transformation rules 
in~\cite{Hama:2010av,Asano:2012gt}. Suppressing spinor indices, the variations of the bosonic fields are 
\bal
\label{SUSY2}
\cQ A_{\mu}&=\frac{i}{k}\xi_{a\dot b}\gamma_\mu(q^a\bar\psi^{\dot b}-\epsilon^{ac}\epsilon^{\dot b\dot c}\psi_{\dot c}\bar q_{c} -\bar{\tilde q}^{\,\dot b}\tilde \psi^{a}
+\epsilon^{\dot b\dot c}\epsilon^{ac}\bar{\tilde\psi}_{c}\tilde q_{\dot c})\,,
\hskip-7cm\\
\cQ q^a&=\xi^{a\dot b}\psi_{\dot b}\,,
\qquad&
\cQ\bar q_{a}&=\xi_{a\dot b}\bar\psi^{\dot b}\,,
\\
\cQ \tilde q_{\dot b}&=-\xi_{a\dot b}\tilde\psi^{a}\,,
\qquad&
\cQ\bar{\tilde q}^{\dot b}&=-\xi^{a\dot b}\bar{\tilde\psi}_{a}\,,
\hskip5cm
\eal
This is all in Euclidean signature and in the frame outlined below $\gamma_\varphi=\sigma_3$ .

The supersymmetry variation of the connection \eqref{eqn:cL} is then
\bal
\label{susy-var}
\cQ \cL &= \frac{i}{k}\left( \xi^\beta_{a \dot a} (\sigma_3)\indices{_\beta^\alpha} 
+ M\indices{_a^b} \xi_{b \dot a}^\alpha \right) q^a \bar\psi^{\dot a}_{\alpha}
- \frac{i}{k}\left( \xi^\beta_{a \dot a} (\sigma_3)\indices{_\beta^\alpha} 
- \xi_{b \dot a}^\alpha \epsilon^{bc}M\indices{_c^d}\epsilon_{da}   \right)\psi^{\dot a}_\alpha \bar q^a \\
&\quad - \frac{i}{k}\left( \xi^\beta_{a \dot a} (\sigma_3)\indices{_\beta^\alpha} 
+ \widetilde M\indices{_{\dot a}^{\dot b}} \xi_{a \dot b}^\alpha \right)
{\bar{\tilde q}}^{\dot a} \tilde \psi^a_\alpha 
+ \frac{i}{k}\left( \xi^\beta_{a \dot a} (\sigma_3)\indices{_\beta^\alpha} 
- \xi_{a \dot b}^\alpha \epsilon^{{\dot b}{\dot c}} 
\widetilde M\indices{_{\dot c}^{\dot d}}\epsilon_{{\dot d}{\dot a}} \right) 
{\bar{\tilde \psi}^a_\alpha \tilde q^{\dot a}} \,. 
\eal
The supercharge $\cQ$ is a linear combination of the 16 supercharges $Q^{a\dot a}_l$, $Q^{a\dot a}_r$, 
$Q^{a\dot a}_{\bar l}$ and $Q^{a\dot a}_{\bar r}$ given by ($\imath$ takes values $l$, $r$ and likewise $\bar\imath$)
\beq
\label{cQ}
\cQ=\eta_{a \dot{a}}^{\imath} Q_{\imath}^{a\dot{a}}+ \bar\eta_{a \dot{a}}^{\imath} (\sigma_1)\indices{_\imath^{\bar\jmath}} Q_{\bar{\jmath}}^{a\dot{a}}\,.
\eeq
The interpolating $\sigma_1$ guarantees that $\eta^\imath_{a \dot a}, \bar\eta^{\imath}_{a\dot a}$ transform in the same representation of the conformal group, see Appendix~\ref{app:symmetries} for details. 
The right-hand-side of \eqref{susy-var} is expressed in terms of $\xi^{\alpha}_{a \dot{a}}$, which package together the 
$\eta$ parameters and the four Killing spinors $\xi^\imath_\alpha$, $\xi^{\bar\imath}_\alpha$
\beq
\label{eqn:relationlr}
\xi_{a\dot{a}}^{\alpha}= \eta_{a\dot{a}}^{\imath} \xi_{\imath}^{\alpha} + \bar\eta_{a\dot{a}}^{\imath} (\sigma_1)\indices{_\imath^{\bar\imath}} \xi^\alpha_{\bar{\imath}}\,.
\eeq
Note that the Killing spinors appear here with raised spinor and lowered $\imath$, $\bar\imath$ indices 
compared to \eqref{killingspinors}.

Let us recall some facts about the Killing spinors. They obey the equations
\beq
\label{xi}
\nabla_\mu \xi^{l,\bar l} = \frac{i}{2}\gamma_\mu \xi^{l, \bar l}\,,
\qquad
\nabla_\mu \xi^{r, \bar r} = - \frac{i}{2}\gamma_\mu \xi^{r, \bar r}\,.
\eeq
Specifically, following~\cite{Kapustin:2009kz, Assel:2015oxa}, we can use the Lie group structure of $S^3$ to 
construct a left-invariant dreibein $e_i$ with spin connection $\omega_{ij}=\epsilon_{ijk}e^k$. 
Then the 
spin connection in the first equation cancels the right-hand-side and the solutions are simply constant spinors. 
For the great circle along the $\mu=\varphi$ direction, 
those can be chosen as eigenstates of $\sigma_3$. In this setup, the second equation reads
\beq
\partial_\varphi \xi^{r, \bar r} = -i\sigma_3 \xi^{r, \bar r}\,.
\eeq
Clearly they can again be chosen along the circle to be eigenvectors of $\sigma_3$ such that 
we find
\beq
\label{killingspinors}
\xi^l_\alpha=\begin{pmatrix}1\\0\end{pmatrix},
\qquad
\xi^{\bar l}_\alpha=\begin{pmatrix}0\\1\end{pmatrix},
\qquad
\xi^r_\alpha=\begin{pmatrix}e^{-i\varphi}\\0\end{pmatrix},
\qquad
\xi^{\bar r}_\alpha=\begin{pmatrix}0\\e^{i\varphi}\end{pmatrix}.
\eeq
In particular note that the unbarred spinors have only $+$ components and the barred ones only 
$-$. This makes the indices redundant and allows us to eliminate some of them as already done in \eqref{cQ}. 
With those expressions for the Killing spinors,~\eqref{eqn:relationlr} becomes
\bal
\xi^+_{a\dot a} = \bar\eta^l_{a\dot a} - \bar\eta^r_{a\dot a} e^{i\varphi}, \qquad \xi^-_{a\dot a} = \eta^l_{a\dot a} e^{-i\varphi} - \eta_{a\dot a}^r.
\eal

For the supersymmetry variation of the connection to vanish, all four terms in \eqref{susy-var} must vanish. A Wilson loop can 
also be invariant under a symmetry when the variation of the connection is an appropriate 
covariant derivative~\cite{Drukker:2019bev}, but as there are no derivatives on the right-hand-side, this 
is not the case here. Multiplying from the left the second term in \eqref{susy-var} 
by $M$ and the fourth by $\widetilde M$, and both by 
$\sigma_3$ from the right, we find
\bal
\label{eqn:SUSYCondMatrix}
\xi_{a \dot{a}}^{\beta} (\sigma_3)\indices{_\beta^\alpha} + M\indices{_a^b} \xi_{b \dot{a}}^{\alpha}&=0\,, 
&\qquad
\xi_{a \dot{a}}^{\beta} (\sigma_3)\indices{_\beta^\alpha} + \widetilde M\indices{_{\dot{a}}^{\dot{b}} }\xi_{a \dot{b}}^{\alpha} &=0\,, 
\\
M\indices{_a^b} \xi_{b \dot{a}}^{\alpha} - \det(M) \xi_{a \dot{a}}^{\beta} (\sigma_3)\indices{_\beta^\alpha}&=0\,, 
&\qquad
\widetilde M\indices{_{\dot{a}}^{\dot{b}} } \xi_{a \dot{b}}^{\alpha}  
- \det(\widetilde M) \xi_{a \dot{a}}^{\beta} (\sigma_3)\indices{_\beta^\alpha}&=0\,.
\eal
Comparing the two lines, this can only be solved by all $\xi^\alpha_{a\dot a}=0$ (which means no supersymmetry) or by
\beq
\label{detM}
\det M = \det \widetilde M=-1\, .
\eeq
The supersymmetry parameters then have to satisfy the eigenvector equations
\beq
\label{eqn:SUSYEigenvector}
M\indices{_a^b} \xi^\pm_{b\dot a}=\mp\xi^\pm_{a\dot a}\,,
\qquad
\widetilde M\indices{_{\dot a}^{\dot b}} \xi^\pm_{a\dot b}=\mp\xi^\pm_{a\dot a}\,.
\eeq
If this is solved by any nonzero $\xi^+_{a\dot a}$, both $M$ and $\widetilde M$ must have an eigenvalue 
$-1$, and if it is solved by $\xi^-_{a\dot a}$, one eigenvalue of both $M$, $\widetilde M$ must be $1$. From~\eqref{detM} 
we see that regardless, $M$ and $\widetilde M$ have both the eigenvalues $1$ and $-1$.

The simplest possibility is of course when $M=\widetilde{M}=\diag(1,-1)$. Plugging them back into \eqref{eqn:cL}, 
we find the Gaiotto-Yin loop \cite{Gaiotto:2007qi} with four preserved supercharges 
$Q^{\dot{1}1}_{l,r},Q^{\dot{2}2}_{\bar{l},\bar{r}}$
\beq
\label{eqn:GY}
\cL=A_{\varphi} +\frac{i}{k} (q^1 \bar{q}_1 -q^2 \bar{q}_2 +\bar{\tilde{q}}^{\dot{1}}  \tilde{q}_{\dot{1}} - \bar{\tilde{q}}^{\dot{2}}  \tilde{q}_{\dot{2}})\, .
\eeq
This is indeed the most symmetric and supersymmetric bosonic loop, discussed in more details in Section~\ref{sec:1/4}. 
Up to rotations to other matrices with eigenvalues 1 and $-1$, this is the only possibility with constant matrices. So in fact the 
main focus of this work are the cases when $M$ or $\widetilde M$ have $\varphi$ dependence. For example, 
\beq
\label{lat}
M=\begin{pmatrix}
1 & 0\\
0 & -1
\end{pmatrix},
\qquad
\widetilde{M}=\begin{pmatrix}
\cos \theta & e^{-i\varphi} \sin \theta\\
e^{i\varphi} \sin \theta & -\cos \theta
\end{pmatrix}\, ,
\eeq
also have eigenvalues $1$ and $-1$, but are not constant. Those are ``latitude'' bosonic loops 
\cite{Cardinali:2012ru,Bianchi:2018bke,Drukker:2020dvr}, preserving two supercharges and are discussed in detail in 
Section~\ref{sec:wbarw}. Loops with different values of $\theta$ are not related by symmetry and 
are truly different (and their expectation values are also different \cite{Griguolo:2021rke}). We find 
below one further inequivalent example in the same class as well as a few new classes of Wilson loops 
preserving two supercharges.

To study all the possible supersymmetric bosonic loops systematically, we now proceed to look for the most general 
configurations of $M$ and $\widetilde M$ which allow for nonzero solutions to \eqref{eqn:SUSYEigenvector}. Since 
any $2\times 2$ matrix with two distinct eigenvalues is uniquely determined by its eigenvectors, it is sufficient to determine those. 

Note that in the two equations in~\eqref{eqn:SUSYEigenvector} there is a free parameter. For example
\bal
\xi^-_{a\dot a} = M\indices{_a^b} \xi^-_{b \dot a}\,,
\qquad
\dot a = \dot 1, \dot 2\,.
\eal
In other words, both $\xi^-_{a\dot 1}$ and $\xi^-_{a \dot 2}$ are eigenvectors with the eigenvalue 
$1$, so are linearly dependent. In particular, if we view $\xi^-_{a\dot a}$ as a $2\times2$ matrix, its determinant must 
vanish. This already incorporates the second equation in \eqref{eqn:SUSYEigenvector} and likewise for $\xi^+_{a\dot a}$, 
giving
\bal
\label{eqn:linearDependence}
{\textstyle\det}_{a\dot a}(\xi^-_{a\dot a})
=\frac{1}{2}\epsilon^{ab} \epsilon^{\dot{a}\dot{b}}\xi^-_{a \dot a}\xi^-_{b \dot b}=0\,,
\qquad 
{\textstyle\det}_{a\dot a}(\xi^+_{a\dot a})
=\frac{1}{2} \epsilon^{ab} \epsilon^{\dot{a}\dot{b}}\xi^+_{a \dot a}\xi^+_{b \dot b}=0\,.
\eal

Being of rank $\leq1$, we can clearly write $\xi^\pm_{a\dot a}$ as the outer product of two vectors. But recall that 
these are linear combinations of two Killing spinors
\beq
\xi^-_{a\dot a}=\eta^\imath_{a\dot a}\xi^-_\imath\,.
\eeq
The expression for the determinant translates to
\beq
{\textstyle2\det}_{a\dot a}(\xi^-_{a\dot a})
=\epsilon^{ab} \epsilon^{\dot{a} \dot{b}}\left((\xi^-_l)^2\eta^l_{a\dot a} \eta^l_{b \dot b} 
+2\xi^-_l\xi^-_r\eta^l_{a\dot a} \eta^r_{b\dot b} 
+ (\xi^-_r)^2 \eta^r_{a\dot a} \eta^r_{b\dot b}\right).
\eeq
Since $\xi^-_\imath$ are different functions (as are their squares), \eqref{killingspinors}, this vanishes only if the 
three terms vanish separately
\beq
{\textstyle\det}_{a\dot a}(\eta^l_{a\dot a})
={\textstyle\det}_{a\dot a}(\eta^r_{a\dot a})
=\epsilon^{ab} \epsilon^{\dot{a} \dot{b}}\eta^l_{a\dot a}\eta^r_{b\dot b}
=0\,.
\eeq
The first two equations allow us to represent $\eta$ as a the product of two vectors 
(no sum implied on the right-hand-side)
\beq
\label{eqn:eta}
\eta_{a \dot{a}}^{\imath}=w_a^{\imath} z_{\dot{a}}^{\imath}\,,
\eeq
where $w_a^{\imath}$ and $z_{\dot{a}}^{\imath}$ are constants and the remaining 
condition is
\beq
\label{eqn:CaseConditions0}
(\epsilon^{ab} w_a^l w_b^r)(\epsilon^{\dot{a}\dot{b}} z_{\dot{a}}^l z_{\dot{b}}^r)
={\textstyle\det}_{a\imath}(w_a^\imath){\textstyle\det}_{\dot a\jmath}(z_{\dot a}^\jmath)
={\textstyle\det}_{a\dot a}\bigg(\sum_{\imath}w_a^\imath z_{\dot a}^\imath\bigg)=0\,.
\eeq

All the same arguments carry over to $\xi^+_{a\dot a}$, expressing
\beq
\label{eqn:bareta}
\bar\eta_{a \dot{a}}^{{\imath}}=\bar{w}_a^{{\imath}} \bar{z}_{\dot{a}}^{{\imath}}\,,
\eeq
subject to the constraint
\beq
(\epsilon^{ab} \bar{w}_a^{{l}} \bar{w}_b^{{r}})(\epsilon^{\dot{a}\dot{b}} \bar{z}_{\dot{a}}^{{l}} \bar{z}_{\dot{b}}^{{r}})
={\textstyle\det}_{a\dot a}\bigg(\sum_{\imath}\bar w_a^{\imath}\bar z_{\dot a}^{\imath}\bigg)=0\,.
\eeq

We may define two matrices of Killing spinors $\Xi^-=\diag(\xi^-_l,\xi^-_r)$ and 
$\Xi^+ = \diag(\xi^+_{\bar r}, \xi^+_{\bar l})$, where the indices are properly raised 
and lowered with respect to \eqref{killingspinors}. Also, it is natural to incorporate the action 
of $\sigma_1$, as in \eqref{eqn:relationlr}, such that they both have two unbarred $\imath$ indices and we can now combine 
\eqref{eqn:eta}, \eqref{eqn:bareta} to write $\xi^\pm_{a\dot a}$ as
\beq
\xi_{a \dot{a}}^-=w_a^{\imath}\Xi^-_{\imath\jmath} z_{\dot{a}}^{\jmath}\,,
\qquad
\xi_{a \dot{a}}^+=\bar w_a^{\imath}\Xi^+_{\imath\jmath} \bar z_{\dot{a}}^{\jmath}\,.
\eeq

Going back to \eqref{eqn:CaseConditions0}, there is a nonzero $\xi^-_{a\dot a}$ giving a preserved supersymmetry 
if either ${\textstyle\det}_{a\imath}(w_a^\imath)$ or ${\textstyle\det}_{\dot a\jmath}(z_{\dot a}^\jmath)$ vanish. 
To enumerate the different possibilities:
\begin{enumerate}
\item
\label{factor-w}
$w_a^\imath=w_ay^\imath$, but can immediately absorb $y^\imath$ in $z_{\dot a}^\imath$, 
such that $\eta_{a\dot{a}}^{\imath}=w_a z_{\dot{a}}^{\imath}$.
\item
\label{factor-z}
$z_{\dot a}^\imath$ factors, so 
$\eta_{a\dot{a}}^{\imath}=w_a^{\imath} z_{\dot{a}}$.
\item
\label{factor-both}
If both determinants vanish independently, 
we have $\eta^{\imath}_{a \dot a} = y^{\imath} w_a z_{\dot a}$.
\end{enumerate}
In the first two cases all the components of $\eta^\imath_{a\dot a}$ are related, representing a single 
preserved supercharge. In the last case the factorization to $y^\imath$ means that 
we have two independent solutions with 
$\eta^l_{a\dot a}=w_a z_{\dot a}$ and another one with $\eta^r_{a\dot a}=w_a z_{\dot a}$.

All the same analysis carries over to $\xi^+_{a\dot a}$ allowing to find zero, one or two independent 
solutions.

Lastly, because $\xi^-_{a\dot a}$ and $\xi^+_{a\dot a}$ have different eigenvalues, they cannot be 
proportional to each-other, so for any pair of nonzero eigenvectors
\beq
\label{linearind}
\epsilon^{ab} \xi^-_{a \dot a} \xi^+_{b \dot b} \neq 0\,,
\qquad
\epsilon^{\dot a \dot b} \xi^-_{a \dot a} \xi^+_{b \dot b} \neq 0\,.
\eeq
When both $\xi^-_{a\dot a}$ and $\xi^+_{a\dot a}$ factorise, this translates into conditions such 
as $\epsilon^{ab}w_a\bar w_b\neq0$.

\section{Representative examples}
\label{sec:preresentatives}

We are now ready to write down examples of Wilson loops preserving varying number of supercharges. 
In this section we list the cases up to actions of symmetries explained in Appendix~\ref{app:symmetries}. 
We also outline the proof that the examples we present are indeed representatives 
of every orbit and demonstrate some details of the other elements in the orbit. 

The relevant symmetries 
are the R-symmetry group $SU(2)_L\times SU(2)_R$ (and its complexification) acting on the indices 
$a$ and $\dot a$ 
respectively and the conformal group $SL(2,\bR)$ (and its complexification), acting on functions of 
$\varphi$ (and on the indices $l$, $r$, $\bar l$, $\bar r$). 
In addition the supersymmetry equations \eqref{susy-var} have two discrete transformations 
relating different solutions:
\begin{itemize}
\item
The exchange $M\leftrightarrow\widetilde M$ or equivalently $SU(2)_L\leftrightarrow SU(2)_R$ 
relates cases with factorised $w$ and/or $\bar w$ to cases with factorised $z$ and/or $\bar z$.
\item
The simultaneous change of sign of $M$ and $\widetilde M$ exchanges the $1$ and $-1$ eigenvalues in 
\eqref{eqn:SUSYEigenvector}. Since those are matched to the spinor index and then via \eqref{killingspinors} 
to the exchange $l,r\leftrightarrow\bar l, \bar r$. This has the effect of relating cases with factorised 
$w$ and/or $z$ to cases with factorised $\bar w$ and/or $\bar z$.
\end{itemize}
The symmetry actions are explained in more detail in Appendix~\ref{app:symmetries}.

If we know two eigenvectors of $M$ with the different eigenvalues, say $\xi^-_{a\dot 1}$ and $\xi^+_{a\dot 2}$, 
we can easily reconstruct $M$ as
\beq
\label{constructM}
M\indices{_a^b}=\begin{pmatrix}
\xi^-_{a\dot 1}&\xi^+_{a\dot 2}
\end{pmatrix}
\begin{pmatrix}
1&0\\0&-1
\end{pmatrix}
\begin{pmatrix}
\xi^-_{b\dot 1}&\xi^+_{b\dot 2}
\end{pmatrix}
^{-1}
=\frac{1}{\epsilon ^{ab} \xi_{a\dot{1}}^- \xi_{b\dot{2}}^+} \begin{pmatrix}
\xi_{1\dot{1}}^- \xi_{2\dot{2}}^+ + \xi_{2\dot{1}}^- \xi_{1\dot{2}}^+ & - 2\xi_{1\dot{1}}^- \xi_{1\dot{2}}^+\\
2 \xi_{2\dot{1}}^- \xi_{2\dot{2}}^+ & - \xi_{1\dot{1}}^- \xi_{2\dot{2}}^+ - \xi_{2\dot{1}}^- \xi_{1\dot{2}}^+
\end{pmatrix},
\eeq
and likewise $\widetilde M$. In this way we can find all BPS Wilson loops by going over all possible 
preserved supercharges. If the loop preserves both $\xi^-_{a\dot 1}$ and $\xi^-_{a\dot 2}$ 
(and say one $\xi^+_{a\dot 1}$), we would 
get the same expression from any linear combination of the two $\xi^-$. The cases with single $\xi^-$ or 
$\xi^+$ arise from not fully factorised $\eta^\pm$, which is different from a linear combination of 
factorised ones.

In the case when the loop preserves only one or two $\xi^-$ and no $\xi^+$, 
we can plug into the formula above any vector $\xi^+$ which is not proportional to the eigenvector $\xi^-$ 
to get an appropriate $M$. As $\xi^+$ is any vector not necessarily made of the Killing spinors, it 
can have arbitrary dependence on $\varphi$, and in turn so does $M$. See Section~\ref{sec:wz} 
and Section~\ref{sec:1/16} below.

\subsection{1/4 BPS loops}
\label{sec:1/4}
The most supersymmetric bosonic loops preserve two components of 
$\xi^-_{a\dot a}$ and two of $\xi^+_{a\dot a}$. They are given by 
$\eta^{\imath}_{a \dot a} = y^{\imath} w_a z_{\dot a}$ and
$\bar\eta^{\imath}_{a \dot a} =\bar y^{\imath}\bar w_a \bar z_{\dot a}$. We can choose 
$w_a=\delta_a^1$, $z_{\dot{a}}=\delta_{\dot{a}}^{\dot{1}}$, $\bar{w}_a=\delta_a^2$, 
$\bar{z}_{\dot{a}}=\delta_{\dot{a}}^{\dot{2}}$ such that the four independent supercharges have the 
nonzero parameters
\beq
\label{1/4eta}
\eta^{l}_{1 \dot 1}\,,
\qquad
\eta^{r}_{1 \dot 1}\,,
\qquad
\bar\eta^{l}_{2 \dot 2}\,,
\qquad
\bar\eta^{r}_{2 \dot 2}\,.
\eeq
Clearly they are eigenvector with eigenvalues $\pm1$ of
\beq
M=\widetilde M=\begin{pmatrix}1&0\\0&-1\end{pmatrix},
\eeq
which is also what one gets from \eqref{constructM}, matching \eqref{eqn:GY}.

This is the $\cN=4$ avatar of the Wilson loops first constructed by Gaiotto and Yin~\cite{Gaiotto:2007qi} 
and then rediscovered as the $1/6$ BPS loops of ABJM theory~\cite{Drukker:2008zx, Chen:2008bp, Rey:2008bh}. 
These $M$ and $\widetilde M$ break $SU(2)_L\times SU(2)_R$ to $U(1)^2$. Conjugation of $M$ by elements 
of the complexified $SL(2,\bC)_L$ can produce any other constant matrix with eigenvalues $\pm1$, and likewise 
for $\widetilde M$. So this is the orbit of the broken symmetry group.

Looking at it from the point of view of the parameters $w_a$, $\bar w_a$, $z_{\dot a}$, $\bar z_{\dot a}$, 
$y^\imath$ and $\bar y^{\imath}$, starting with any other factorised choice, we can use two $SL(2,\bC)_{L, R}$
actions to set $w_a=\delta_a^1$ and $z_{\dot{a}}=\delta_{\dot{a}}^{\dot{1}}$ as in the above 
example. This on its own does not fix the barred parameters. But as they cannot be linearly 
dependent on the unbarred ones \eqref{linearind}, we can use the Gram-Schmidt process to produce 
$\bar{w}_a\propto\delta_a^2$, $\bar{z}_{\dot{a}}\propto\delta_{\dot{a}}^{\dot{2}}$, and then can rescale them 
to reproduce \eqref{1/4eta}.

The superalgebra generated by the four preserved supercharges in \eqref{1/4eta} includes the $SO(2,1)_C$ 
conformal group of the circle, guaranteeing that these loops are conformal~\cite{Bianchi:2017ozk, Drukker:2022ywj}. 
In fact, of all the loops discussed in this paper, 
these are the only ones that are conformal and all the other ones have explicit $\varphi$ dependence in 
$M$ and/or $\widetilde M$.

\subsection{3/16 BPS loops}
\label{sec:3/16}
This case arises when three of $w$, $z$, $\bar{w}$, $\bar{z}$ factorise while the remaining one doesn't. 
Let us focus on the case when $\bar z$ does not factorise. As above, we take $w_a=\delta_a^1$, 
$z_{\dot{a}}=\delta_{\dot{a}}^{\dot{1}}$, $\bar{w}_a=\delta_a^2$ and 
$\bar{z}_{\dot a}^{\imath}=\delta^{\imath}_{\dot a}$ (with $l\simeq\dot1$).

The preserved supercharges are then linear combinations of 
\beq
Q_l^{1\dot 1}\,,
\qquad
Q_r^{1\dot 1}\,,
\qquad
Q_{\bar r}^{2\dot 1}
+Q_{\bar l}^{2\dot 2}\,,
\eeq
and the connection is as in \eqref{eqn:cL} with
\beq
\label{M3/16}
M=\begin{pmatrix}
1& 0\\
0& -1
\end{pmatrix},\qquad \widetilde{M}=\begin{pmatrix}
1& 2 e^{-i\varphi}\\
0& -1
\end{pmatrix}.
\eeq
This is a new example of a Wilson loop with an exotic number of preserved supercharges. 
Unlike the 1/4 BPS loops, this is not conformally invariant, as is evident by the explicit 
$\varphi$ dependence in $\widetilde M$. Acting with the broken $SU(2)_L\times SU(2)_R\times SO(2,1)_C$ 
(possibly complexified) gives a rich orbit of further examples. Note that the action of the complexified 
conformal group $SL(2,\bC)$ can transform $e^{i\varphi}$ to any fractional linear function. 
Other examples include the replacement of 
$M\leftrightarrow\widetilde M$ as well as changing their signs. Those correspond to a choice 
of a different parameter among $w$, $z$, $\bar{w}$, $\bar{z}$ that doesn't factorise. 
In particular the example with lower triangular $\widetilde M$ is related by symmetry to 
the case where $M$ and $\widetilde M$ have opposite signs.

To prove that there is a single orbit of the group and \eqref{M3/16} is indeed a representative, 
note that any unfactorised $\bar z$ is a rank 2 matrix which can be brought into the form above with the action of 
either $SL(2,\bC)_R$ or $SL(2,\bC)_C$. It is still invariant under conjugation by any identical element of the 
two groups, which means we have $SL(2,\bC)_R$ freedom to set $z_{\dot{a}}=\delta_{\dot{a}}^{\dot{1}}$. 
Then the $SL(2,\bC)_L$ action can bring $\bar w$ into the desired 
form. This procedure leaves $w$ as a vector not parallel to $\bar w$, so we can choose a linear combination 
producing $w_a=\delta_a^1$.

Supersymmetry enhancement can be easily seen by exploiting the action of $SL(2,\bC)_C$, 
which allows us to apply arbitrary constant rescalings to $e^{i\varphi}$. In the limit 
where the phase in~\eqref{M3/16} vanishes, supersymmetry is enhanced to 1/4 BPS.

\subsection{1/8 BPS loops}
\label{sec:1/8}

There are of course $\binom{4}{2}=6$ different pairs out of $w$, $z$, $\bar{w}$, $\bar{z}$ to factorise, but 
they are pairwise related by extra symmetries. We discuss the three inequivalent classes below.

\subsubsection{Factorised $w$ and $\bar{w}$}
\label{sec:wbarw}

In this case we can take representatives with $w_a=\delta_a^1$, $\bar{w}_a=\delta_a^2$ and 
${z}_{\dot a}^{\imath}=\delta^{\imath}_{\dot a}$. To get to this form from an arbitrary unfactorised 
${z}_{\dot a}^{\imath}$, we can either act from the left with $SL(2,\bC)_R$ or from the right with 
$SL(2,\bC)_C$. This form is still invariant under conjugation by the same elements of the two 
groups, so that is the remaining freedom we have to act on $\bar z$, as well as overall rescaling, 
which is immaterial. By rescaling we can make $\det_{\dot a \imath}\bar z=1$ (since it has 
rank 2) and then by conjugation bring it to Jordan normal form
\beq
\label{almost-lat-z}
\bar{z}_\lambda
=\begin{pmatrix}
\bar z_{\dot{1}}^{ l} & \bar z_{\dot{1}}^{r}\\
 \bar z_{\dot{2}}^{l} &\bar z_{\dot{2}}^{r}  
\end{pmatrix}
=\begin{pmatrix}
\lambda&0\\
0&1/\lambda
\end{pmatrix},
\qquad
\bar{z}'
=\begin{pmatrix}
1&1\\
0&1
\end{pmatrix}.
\eeq

Plugging the diagonal case into \eqref{constructM} gives
\beq
\label{almost-lat}
M=\begin{pmatrix}
1 & 0\\
0 & -1
\end{pmatrix},
\qquad 
\widetilde{M}=\frac{1}{\lambda - \lambda^{-1}} 
\begin{pmatrix}-\lambda - \lambda^{-1} & -2 \lambda e^{-i\varphi} \\ 
2 \lambda^{-1} e^{i\varphi} & \lambda + \lambda^{-1}
\end{pmatrix}.
\eeq

These loops are in fact related to the ``bosonic latitude loops'' \eqref{lat} 
of~\cite{Cardinali:2012ru,Bianchi:2018bke,Drukker:2020dvr}. To see that, 
we take $\cos\theta = - (\lambda + \lambda^{-1})/(\lambda - \lambda^{-1})$, such that
\beq
\widetilde{M}_\theta=
\begin{pmatrix}
\cos\theta& i\lambda e^{-i\varphi}\sin\theta\\
-i\lambda^{-1}e^{i\varphi}\sin\theta & -\cos\theta\,.
\end{pmatrix}.
\eeq
Conformal symmetry acts on $e^{\pm i\varphi}$ as M\"obius transformations, which in particular includes 
the rescaling that eliminates $i\lambda$ from the matrix above, reproducing \eqref{lat}. This can also 
be realised by conjugating $\widetilde M$ by
\beq
\begin{pmatrix}
 1/\sqrt{i\lambda} & 0\\
0 & \sqrt{i\lambda}
\end{pmatrix},
\eeq
which is an $SL(2,\bC)_R$ transformation.

In terms of $z$ and $\bar z$, this $SL(2,\bC)_R$ acts on them from the left giving
\beq
\label{latzzbar}
z
\to\begin{pmatrix}
 1/\sqrt{i\lambda} & 0\\
0 & \sqrt{i\lambda}
\end{pmatrix},
\qquad
\bar{z}_\lambda
\to\begin{pmatrix}
-i\sqrt{i\lambda} & 0\\
0 & i/\sqrt{i\lambda}
\end{pmatrix}.
\eeq
Inserting the resulting $\xi^\pm_{a\dot a}$ into~\eqref{constructM} and taking the above relation between 
$\lambda$ and $\theta$ produces the same result \eqref{lat}. Since the original and new $z$ and $\bar z$ 
are all diagonal, the exact same result can be achieved by right multiplication, which is an $SL(2,\bC)_C$ 
transformation. The fact that we can act with either of the groups indicates that these loops are 
invariant under a particular combination of the two group actions, which is a known symmetry of the 
latitude~\cite{Griguolo:2021rke, Drukker:2022ywj}.

Turning to the non-diagonal case in \eqref{almost-lat-z}. This is a new example, which to our knowledge has 
not been previously described. We find the matrices
\beq
\label{new18}
M=\begin{pmatrix}
1 & 0\\
0 & -1
\end{pmatrix},
\qquad 
\widetilde{M}=\begin{pmatrix}
 -1 + 2 e^{-i\varphi} & -2 e^{-i\varphi} +2 e^{-2i\varphi} \\ -2 & 1 -2e^{-i\varphi} 
\end{pmatrix}.
\eeq
Another representative of this orbit is given by
\beq
\label{newJzzbar}
z = \begin{pmatrix} 1 & -1 \\ 0 & -1 \end{pmatrix}, 
\qquad 
\bar z = \begin{pmatrix} 0 & 1 \\ 1 & 1 \end{pmatrix},
\eeq
which leads to
\beq
\label{M18nondiag}
M=\begin{pmatrix}
1 & 0\\
0 & -1
\end{pmatrix},
\qquad
\widetilde{M} = \begin{pmatrix}
1 - 2 e^{2i\varphi} & 2 e^{i\varphi} \left( 1 + e^{i\varphi} \right) \\
2 e^{i\varphi} \left( 1 - e^{i\varphi} \right) & -1 + 2 e^{2i\varphi}
\end{pmatrix}.
\eeq
Instead of constructing the detailed map between the two cases, as we did for the latitudes, note that 
$z$ and $\bar z$ offer a simple way to identify the orbit. Since $z$, $\bar z$ both transform in the fundamental 
of $SL(2,\bC)_R$ and the antifundamental of $SL(2,\bC)_C$. 
The matrix $z^{-1}\bar z$ transforms in the adjoint 
of $SL(2,\bC)_C$ and is invariant under the other group. Conversely $\bar zz^{-1}$ is in the adjoint of 
$SL(2,\bC)_R$ and invariant under the other group. The eigenvalues of these matrices are then invariant 
under both groups, but these two matrices have the same eigenvalues. Since any nonzero rescaling is immaterial, 
we can always set them to $\lambda$ and $1/\lambda$ and compare with \eqref{almost-lat-z}.

Clearly for the latitudes in \eqref{latzzbar}, we reproduce the same eigenvalues as $\bar z_\lambda$ in \eqref{almost-lat-z}. 
Likewise, the Jordan form of those for \eqref{newJzzbar} is the same as $z'$.

The case with factorized $z$ and $\bar z$ gives similar loops under the exchange $M
\leftrightarrow\widetilde M$.

To see how to get supersymmetry enhancement to the previous examples, the case of 
enhancement to 4 supercharges in Section~\ref{sec:1/4} is obvious, by taking 
$\theta\to0$ in \eqref{lat}.

To get to the 3/16 BPS case in Section~\ref{sec:3/16}, we need to take 
$\lambda\to\infty$ in \eqref{almost-lat}. In terms of $\theta$, this is a double 
scaling limit by first 
using a complexified conformal transformation that scales $e^{i\varphi}\to 2e^{i\varphi}/\theta$ 
and $e^{-i\varphi}\to \theta\,e^{-i\varphi}/2$ and then take $\theta\to0$. The expression in \eqref{lat} 
then clearly becomes the transpose of \eqref{M3/16}, which is another example 
of a 3/16 BPS loop.

The nondiagonalisable case admits similar limits. Rescaling $e^{-i\varphi}$ in~\eqref{new18} 
allows us to tune out the $\varphi$ dependent terms entirely, which brings us back to the 1/4 BPS case. To hit a 
3/16 BPS orbit instead, we first rescale the phases by $x^2$, and then conjugate 
$\widetilde{M}$ with $\diag(x, x^{-1})$, which 
shifts one factor of $x^2$ from the bottom left to the top right. The limit $x \to 0$ 
then removes all but one phase, and we recover the 3/16 BPS loop~\eqref{M3/16}, as 
before.

\subsubsection{Factorised $w$ and $\bar{z}$}
\label{sec:wbarz}
As an illustration of this case, we can take $w_a= \delta_a^1$, $\bar{z}_{\dot{a}}=\delta_{\dot{a}}^{\dot{2}}$, 
$\bar{w}_a^{\imath}= \delta_a^1 \delta^\imath_l - \delta_a^2 \delta^\imath_r$ and ${z}_{\dot a}^{\imath}=\delta^{\imath}_{\dot a}$ 
to get
\beq
\label{mixed18}
M=\begin{pmatrix}
1 & -2e^{-i\varphi}\\
0 & -1
\end{pmatrix},
\qquad
\widetilde{M}=\begin{pmatrix}
1 & 0\\
-2e^{i\varphi} & -1
\end{pmatrix}.
\eeq

Unlike the previous example, here we could use the symmetry to choose a unique representative, so there 
is only one conjugacy class (the argument follows the same logic as in the previous examples). 
To our knowledge, such loops have not been previously described. The action of the conformal 
group on these loops produces more loops with fractional linear functions in both $M$ and $\widetilde M$.

The case when $\bar w$ and $z$ are instead factorised is related again by $M\leftrightarrow\widetilde M$.

The loop~\eqref{mixed18} admits 3/16 BPS limits. Conjugation with $\diag(x, x^{-1}) \in SL(2,\bC)_L$ 
and taking $x\to\infty$ allows us to tune out the phase in $M$ and the same can be done with $\widetilde M$.

\subsubsection{Factorised $w$ and $z$}
\label{sec:wz}

When $w$ and $z$ both factorise, we have two preserved $\xi^-$ supercharges and no $\xi^+$ ones. We choose 
representative supercharges with $w_a=\delta_a^1$ and $z_{\dot{a}}=\delta_{\dot{a}}^{\dot{1}}$. This does not 
completely fix $M$ and $\widetilde M$, as \eqref{constructM} requires also to specify $\xi^+$. From the above information 
alone, we find
\beq
M\indices{_1^1}=\widetilde{M}\indices{_1^1}=1\,,
\qquad 
M\indices{_2^1}=\widetilde{M}\indices{_2^1}=0\,.
\eeq
Then, since $\det M=\det \widetilde{M}=-1$, we get $M\indices{_2^2}=\widetilde{M}\indices{_2^2}=-1$, which 
leaves $M\indices{_1^2}$ and $\widetilde{M}\indices{_1^2}$ as completely arbitrary periodic functions of $\varphi$. 
We denote them $2 \mu(\varphi)$ and $2 \tilde{\mu}(\varphi)$ respectively to get 
\beq
M=\begin{pmatrix}
1 & 2 \mu(\varphi)\\
0 & -1
\end{pmatrix},
\qquad 
\widetilde{M}=\begin{pmatrix}
1 & 2 \tilde{\mu}(\varphi)\\
0 & -1
\end{pmatrix}.
\eeq
Alternatively, we can arrive at the same result by choosing the second eigenvectors for $M$ and 
$\widetilde M$ as
\beq
\begin{pmatrix} \mu(\varphi) \\ -1 \end{pmatrix},
\qquad
\begin{pmatrix} \tilde\mu(\varphi) \\ -1 \end{pmatrix}.
\eeq
Of course, if $\mu(\varphi)$ or $\tilde\mu(\varphi)$ are constants or $e^{\pm i\varphi}$, these eigenvectors 
are in fact Killing spinors and the loop will have enhanced supersymmetry to $3/16$ or $1/4$ and will 
match the forms in Section~\ref{sec:3/16} or Section~\ref{sec:1/4} up to symmetry action.

To understand the reason for this freedom of arbitrary functions in $M$ and $\widetilde M$, it is instructive to 
reconstruct the general supercharge preserved by these loops, that is \eqref{cQ} 
\beq
\cQ=\eta^{\imath}_{1 \dot 1}Q_\imath^{1\dot 1}\,.
\eeq
Examining the supersymmetry variations \eqref{SUSY2}, we clearly see that $\cQ (q^1\bar q_2)=\cQ (\bar{\tilde q}^{\dot1}\tilde q_{\dot2})=0$. So this Wilson loop 
is simply the 1/4 BPS Wilson loop of Section~\ref{sec:1/4} with arbitrary insertions of these 
bilinears that are chiral under this pair of supercharges.

\subsection{1/16 BPS loops}
\label{sec:1/16}
Among the four possible cases, we choose the one with factorised $w$, so in particular no $\xi^+$ supercharges. 
If we choose $w_a=\delta_a^1$ and ${z}_{\dot a}^{\imath}=\delta^{\imath}_{\dot a}$, then for $M$ things are similar 
to the last case, where we get $M\indices{_1^1}=1$, $M\indices{_2^1}=0$, and by $\det M=1$ we have the other two entries 
$M\indices{_1^2}=2\mu(\varphi)$, $M\indices{_2^2}=-1$. The second matrix, $\widetilde{M}$, is different. 
Its entries satisfy the two equations
\beq
\widetilde{M}\indices{_{\dot{1}}^{\dot{1}}} e^{-i\varphi} -\widetilde{M}\indices{_{\dot{1}}^{\dot{2}}}
=e^{-i\varphi}\,,
\qquad 
\widetilde{M}\indices{_{\dot{2}}^{\dot{1}}} e^{-i\varphi} -\widetilde{M}\indices{_{\dot{2}}^{\dot{2}}} =-1\,,
\eeq
and there is the extra condition $\det \widetilde{M}=-1$, $\tr\widetilde{M} = 0$. It therefore still has one completely 
free parameter and we take $\widetilde{M}\indices{_{\dot{1}}^{\dot{1}}}=\tilde\mu(\varphi)$. Then
\beq
M=\begin{pmatrix}
1 & 2\mu(\varphi)\\
0 & -1
\end{pmatrix},
\qquad
\widetilde{M}=\begin{pmatrix}
\tilde\mu(\varphi) & e^{-i\varphi}(\tilde\mu(\varphi)-1)\\
-e^{i\varphi} (\tilde\mu(\varphi)+1) & -\tilde\mu(\varphi)
\end{pmatrix}.
\eeq

\section{Theories without twisted hypers}
\label{sec:without}

In the discussion so far, we assumed that the theory contains both hypermultiplets and twisted hypermultiplets, 
but this is not necessary for $\cN=4$ supersymmetry. The discussion follows through if the theory has only hypermultiplets. 
The supersymmetry variations are as in \eqref{SUSY2} except that we should remove all the twisted 
hypermultiplet fields.

The ansatz for the Wilson loops is as in \eqref{eqn:cL}, but without the term involving $\widetilde M$. The theory has the same 16 
supercharges and Killing spinors, but the supersymmetry conditions impose far fewer constraints. Specifically, only the left 
equation in \eqref{eqn:SUSYEigenvector} remains, meaning that there is no constraint on the dotted indices of 
the supercharges.

In terms of the conditions on $\eta$ and $\bar\eta$, the $z$ and $\bar z$ parameters are always factorised, 
so equation \eqref{eqn:eta} becomes
\beq
\eta_{a \dot{a}}^{\imath}=w_a^{\imath} z_{\dot{a}}\,,
\eeq
and the requirement for supersymmetry is that $\det_{a\imath}w_a^{\imath}=0$ or 
$\det_{a\imath}\bar w_a^{\imath}=0$, meaning one of them also factorises.

The completely factorised case is as in Section~\ref{sec:1/4}, with
\beq
M=\begin{pmatrix}1&0\\0&-1\end{pmatrix}.
\eeq
Now this loop preserves 8 supercharges, so is 1/2 BPS.

When only $w$ factorises but not $\bar w$ we have loops similar to the construction in Section~\ref{sec:3/16}
\beq
M=\begin{pmatrix}
1 & 2e^{-i\varphi}\\
0 & -1
\end{pmatrix},
\eeq
which are now $1/4$ BPS. Likewise, when only $\bar w$ factorises we can get a similar $M$ with the phase 
in the bottom left.

In a linear quiver theory with two nodes, bifundamental hypermultiplets and no twisted hypermultiplets, 
there are then two bosonic 
1/2 BPS loops, one at each of the nodes as well as fermionic 1/2 BPS loops. The bosonic loops 
break $SU(2)_L$ and preserve $SU(2)_R$. The 1/2 BPS fermionic loops break $SU(2)_R$ and 
preserve $SU(2)_L$. 

\section*{Acknowledgements}

We would like to acknowledge fruitful discussions with E. Pomoni. 
ND is supported by STFC under the grants  ST/T000759/1 and ST/P000258/1 
and by the National Science Foundation under Grant No. NSF PHY-1748958. 
The work of ZK is supported by CSC grant No. 201906340174.
ND would like to thank \'Ecole Polytechnique F\'ed\'erale de Lausanne, 
the Simons Center for Geometry and Physics, Stony Brook and the KITP, Santa Barbara 
for their hospitality in the course of this work. ZK would like to thank 
DT and the University of Modena and Reggio Emilia as well as T. Fiol and the University of Barcelona 
for their hospitality.


\appendix

\section{Symmetries}
\label{app:symmetries}

The theories studied here have $SO(4) \cong(SU(2)_L \times SU(2)_R)/\bZ_2$ R-symmetry and 
$SO(4,1)$ conformal symmetry, which are packaged together into an $OSp(4|4)$ supergroup. 
The geometry of the circle breaks the conformal group to $SO(2,1)_C\times SO(2)$, and the particular 
choice of line operator can break the symmetry further. However, the bosonic loops all are invariant under 
the transverse $SO(2)$, so what we focus on is the action of $SU(2)_L \times SU(2)_R\times SO(2,1)_C$ 
on the loops constructed in the body of the paper.

To be precise, we are not imposing any reality or hermiticity conditions, so we allow for the action of 
the complexified group $SL(2,\bC)_L\times SL(2,\bC)_R\times SL(2,\bC)_C$. The scalars $q_a$ are 
in the fundamental of $SU(2)_L$ and $\bar q^a$ in the anti-fundamental. The scalars from the twisted 
hyper are charged under $SU(2)_R$. Clearly the matrices $M$ and $\widetilde M$ are in the adjoint 
of $SU(2)_L$ and $SU(2)_R$ respectively. The action on the supercharges as well as on $\eta, \bar\eta$ and 
their decomposition into the parameters $w$, $z$, $y$, $\bar w$, $\bar z$ and $\bar y$ can be read 
off from their index structure.

The action of the conformal group is more involved. The parameters $\eta^\imath_{a\dot a}$ and 
$\bar\eta^{\imath}_{a\dot a}$ are doublets, with the 
indices $\imath$ interchanged under the action of the conformal generators $J_\pm$ satisfying 
the algebra
\beq
[J_0,J_\pm]=\pm J_\pm\,,\qquad
[J_+,J_-]=2J_0\,.
\eeq
Under a finite conformal transformation $U \in  SL(2,\bC)_C$, they transform as
\bal
\eta_{a\dot a}^\iota \longmapsto \eta_{a\dot a}^\jmath \left( U^T \right)\indices{_\jmath^\imath}, \qquad
\bar\eta_{a\dot a}^\iota \longmapsto \bar\eta_{a\dot a}^\jmath \left( U^T \right)\indices{_\jmath^\imath}.
\eal

The action on $M$ and $\widetilde M$ is set by their $\varphi$ dependence. The 1/4 BPS loops in 
Section~\ref{sec:1/4} are all invariant. In all the other cases we know that the conformal group acts on the 
unit circle via M\"obius transformations
\beq
U:\ e^{i\varphi} \ \longmapsto\ \frac{d e^{i\varphi}-b}{-c e^{i\varphi}+a}\,,
\qquad
U=\begin{pmatrix}a&b\\c&d\end{pmatrix},
\qquad
ad-bc=1\,.
\eeq
This determines the transformations of all the Wilson loops discussed in Section~\ref{sec:preresentatives}.

\bibliographystyle{utphys2}
\bibliography{refs}
\end{document}